\documentclass[12pt,preprint]{aastex}

\begin{document}

\title{\bf CH Cyg X-ray Jet Activity and Multi-component Structures}

\author{Margarita Karovska\altaffilmark{1}}
\author{Christopher L. Carilli\altaffilmark{2}}
\author{John C. Raymond\altaffilmark{1}}
\author{Janet A. Mattei\altaffilmark{3}}
\altaffiltext{1}{Smithsonian Astrophysical Observatory, 60 Garden St., 
  Cambridge, MA 02138; mkarovska@cfa.harvard.edu.}
\altaffiltext{2}{National
               Radio Astronomy Observatory, PO Box O, Socorro, NM 87801}
\altaffiltext{3}{American Association of Variable Star Observers, 25 Birch
               Street, Cambridge, MA 02138}

\begin{abstract}

In this paper we report detection of multiple component structures
in a Chandra 
X-ray image obtained in March 2001 of the nearby symbiotic interacting
binary system CH Cyg.
These components 
include a compact central
source, an arc-like
structure or a loop extending to  1.5'' (400 AU) from the central
source associated with
the 1997 jet activity, 
and possibly a
newly formed jet extending to $\sim$ 150 AU from the central source.
The structures are also visible in
VLA and HST images obtained close in time to the Chandra
observations. 
The emission from the loop
is consistent with optically thin thermal X-ray
emission originating from a shock resulting from interaction of the jet ejecta with
the dense circumbinary material. 
The emission from the central source originates within $<$ 50 AU region, and is likely associated with the accretion disk around the
white dwarf.
CH Cyg is only the second symbiotic system
with jet activity detected at X-ray wavelengths, and the
Chandra high-angular resolution image,
combined with the VLA and HST images, 
provides the closest view of the region of jet
formation and interaction with the circumbinary material
in a symbiotic binary.

\end{abstract}

\keywords{accretion, accretion disks --- binaries: close --- stars:
  individual (CH Cyg) --- binaries: symbiotic, stars: winds, outflows --- x-rays: general}

\section{INTRODUCTION}

Symbiotic systems are interacting binaries 
showing a composite spectrum with signatures of 
a late-type giant and a high-temperature component, often a compact
accreting object. As with many other astronomical sources, accretion is thought to be the
primary energy source in these systems.
The interaction between the components occurs via wind accretion
and/or Roche lobe overflow.
Symbiotic systems are a very important subgroup of interacting
binaries because they are among the likely candidates for progenitors
of bipolar planetary nebulae, and they have also been invoked as potential 
progenitors of at least a fraction of Supernovae type Ia, key 
cosmological distance indicators (e.g. Chugai and Yungelson 2004).
So far there are only a couple hundred known members in this class.

CH Cyg is one of few symbiotic systems 
showing jet activity (e.g. Corradi {\it et al.} 2001, Kellogg {\it et al.} 2001).
At a distance of 245$^{+}_{-}$50 pc (Perryman {\it et al.} 1997)
CH Cyg is one of the closest symbiotic systems, therefore providing a
unique opportunity for
high-angular resolution imaging of a close circumbinary environment in
an interacting binary, 
including the region of jet formation and early propagation.

The CH Cyg system is composed of an evolved M6-7 III giant and
an accreting white dwarf.
There is also evidence for a third 
body in the system, also an evolved giant
(e.g. Hinkle {\it et al.} 1993, Scopal {\it et al.} 2002). 
This binary is one of 
the most fascinating objects in its class because of its extremely complex 
circumbinary environment, and its dramatic outbursts.

In the past few decades CH Cyg has undergone several outbursts,
each preceded by extended intervals of quiescence (Karovska and
Mattei 1992, Sokoloski and Kenyon 2003,
Eyres {\it et al.} 2002, Skopal {\it et al.} 2002). 
Since 1984, when a powerful jet was formed in the midst of a prominent radio and optical
outburst (Taylor {\it et al.} 1986), several outbursts have been detected including
in 1996 and in 1998 (Karovska {\it et al.}
1998a\&b, Eyres {\it et al.}
2002).
Radio monitoring of the system has shown indications of a
possible precessing jet with enhanced activity associated with occasional
outbursts (Crocker {\it et al.} 2002).

Observations of CH Cyg have been carried out over the past
two decades using several X-ray satellites.
Spectral modeling indicates a multi-component
structure in the emission from this system
(Leahy and Taylor
1987; Leahy and Volk 1995; Ezuka {\it et al.} 1998; Mukai {\it et
al.} 2006). 
For example, Ezuka {\it et al.} (1998) detected a soft component in
the ASCA spectrum with
 flux of 3.7$\times$ 10$^{-12}$ $ergs$ $cm^{-2}$$s^{-1}$ from 0.5 to 2.5
keV. They modeled this emission as
two optically thin thermal plasma emissions, the temperatures of which are 0.2 and
0.7 keV. However, the origin of the soft component remained unclear.
A previously unknown hard component 
was also detected in the spectrum with flux of 6.9$\times$ 10$^{-12}$ $ergs$ $cm^{-2}$$s^{-1}$ in the region from 2-10 keV.  
This component was modeled with optically thin thermal plasma emission with temperature of 7.3 keV,
undergoing a strong photoelectric absorption with N$_H$ =$\sim$ 10$^{23}$ $cm^{-2}$. 
In addition, several emission features
were detected including
iron emission lines between 6 and 7 keV.
Wheatley and Kallman (2006) re-analyzed the ASCA
spectrum and concluded that there is only one emission
component. They suggest that similar to Seyfert 2 galaxies,
the soft emission is due to scattering of
the hard component in a photoionized medium surrounding the white
dwarf.  Mukai {\it et al.} (2006) analyzed recent Suzaku 
observations and found that the large equivalent width of the 
6.4 keV fluorescence line implies that the hard X-rays are 
indeed dominated by scattering.  However, they concluded that there are two distinct
components responsible for the hard and soft X-ray emission.
They also confirmed that there is a significant long-term variability
in the X-ray emission in the system.

Chandra observations carried out in 2001
detected an elongation in the CH Cyg image, extending 2-4'' to 
the south from the central source. Galloway and Sokoloski (2004)
attributed this extension to jet activity associated
with the white dwarf. This is only 
the second source from a population of few hundred known
symbiotics showing jet activity in X-rays.
However,
in the Galloway and Sokoloski (2004) analysis of the Chandra image the potential spatial resolution of $<$0.5''
on-axis 
was significantly degraded by gaussian smoothing used
to enhance the faint extended emission.
Therefore, the resolution was
insufficient to study directly the spatial distribution of the
X-ray  emission.

In this paper we present results from our further detailed spatial
analysis of the Chandra observation of CH Cyg
combined with the results from
VLA and HST imaging of the system.  Using the
full angular resolution potential of Chandra we identify several
distinct spatial components contributing to the X-ray emission and 
resolve the ambiguity
left from previous X-ray observations regarding the
extended emission and the 
contributors to the soft and the hard component of the X-ray
spectrum of CH Cyg.

\section{CHANDRA OBSERVATIONS AND ANALYSIS}

The Chandra pointed  observations of CH Cyg were carried out over 
47 ks on 2001 March 27 (OBSID 1904) using the ACIS-S instrument in
combination with HETG grating (Weisskopf {\it et al.} 2002).
Significant signal was detected mostly in the
zeroth-order of the spectrum;
a total of 770 counts were detected form 0.2 to 8 keV.
We analyzed the {\it Chandra} observations using {\it CIAO} data
reduction and analysis routines. In our analyses we used Ciao 3 and cal database
2.2 (similar to
Galloway and Sokoloski 2004), as well as  the more recent 
Ciao 3.3 version.
\footnote{CIAO is the Chandra Interactive Analysis of Observation's
data analyses system package (http://cxc.harvard.edu/ciao)}

{\it Chandra} data include information about the photon energies and
positions. This information was used to obtain energy-filtered images
and to carry out
sub-pixel resolution spatial analysis ({\it vis} Karovska {\it et
al.} 2005). 
Based on the three spectral
emission components identified by Ezuka {\it et al.} (1998) and Galloway and
Sokoloski (2004)
(see for example Fig. 3 in Galloway and Sokoloski 2004)
we created images in the soft
band (0.2-2keV), the
hard band  (2-6 kev), and in the spectral region around the strong
emission feature at 6.5-7 attributed to 6.67 keV iron line.
The total counts detected in these images using a circular region with
a radius of 5'' centered on the brightest central source are: 290,
260, and 100 counts, respectively.
The expected X-ray background in the region is negligible.

We explored the spatial
distribution of the emission in the images of the system at 0.2'' 
resolution using deconvolution and smoothing 
techniques.
The pixel size of the Chandra/ACIS detector is 0.492".
However, due to the telescope dither smaller 
spatial scales are accessible as the image moves across the detector pixels.
This sampling of smaller than detector pixel scales allows sub-pixel 
analysis and deconvolution.
Similar techniques were applied for resolving Mira AB 0.5" binary
system (Karovska {\it et al.} 2005),
as well as for SN1987A (e.g. Burrows {\it et al.} 2000, Park {\it et al.} 2002) 
which allowed resolving sub-arcsecond structures
in the 1" scale SN remnant.
After processing of the data (including with pixel randomization turned off)
we binned the images with 0.1'' pixels.

We used PSF simulations for comparison with the observations in order
to evaluate detected structures in
the observed images as well as to estimate the extent of the central source
in different spectral bands.
The PSF
simulations were  carried out using information on the spectral
distribution and off-axis location of the system as inputs to ChaRT PSF
simulator\footnote{http://asc.harvard.edu/chart/index.html}
(Karovska {\it et al.} 2003).

To detect the low contrast diffuse emission we applied the adaptive
smoothing CIAO tool {\it csmooth}, based on the algorithm
developed by Ebeling {\it et al.} (2006) and implemented in {\it asmooth}. 
{\it Csmooth} is an adaptive smoothing tool for
images containing multiscale complex structures,
and it preserves the spatial signatures
and the associated counts as well as significance estimates.
The smoothing is achieved by convolution with a 
gaussian kernel. The size of the kernel increases from a small initial
value and the smoothing scale is increased 
until the total number of counts under the kernel exceeds a 
value that is determined from a preset significance and the 
expected number of background counts in the kernel area. 
Significance is computed at each pixel location 
for each smoothing scale by 
comparing the total counts under the kernel to the 
background in the same area.
Pixels for which the significance exceeds 
the threshold value set by the user are smoothed at the 
current scale, and are excluded from subsequent smoothing 
steps at larger scales.

In Figure 1 we display the smoothed images of CH Cyg in 
the soft (0.2-2), and the
hard (2-6 keV) bands.
In these images we show structures smoothed above 3 $\sigma$ 
significance level, and the smoothing scales vary from 0.2 to 0.5 arcseconds.
A compact central source is detected in all three bands. 
Our analysis
shows that the central sources in the hard band image and in the 6.5 to 7 keV image are consistent with unresolved 
point sources (FWHM less then 0.2'').
In addition to a central source, several extended
structures are detected that are especially prominent in the soft
image.
For the central source,
within a circular region with 0.3'' radius,
we estimate a hardness ratio of $\sim$2,
(counts $>$2keV divided by the counts $<$ 2 keV).

Figure 2 shows the smoothed soft image, as in Figure 1, with overplotted
contours of the simulated Chandra PSF.
When compared to the PSF this image shows a somewhat extended central
source to $\sim$ 0.7'' at P.A. $\sim$150/330 degrees, and an extended 
emission to the south
consisting of at least two components:
a loop-like structure, and
a faint emission extending beyond the loop.
Figure 3 displays the soft
image after subtracting the PSF, showing the extended structure in
the central source at P.A. $\sim$ 330 degrees and the loop-like 
structure to the south.

The loop-like structure
 extends to
a distance of
$\sim$1.5'' from the central source. 
The southern part of the loop appears brighter and resembles an arc.
There are regions of enhanced brightness (3-5 $\sigma$) in the
southern portion of this loop, smoothed on scales of 0.3 arcseconds.
We detect $\sim$40 counts in the region of the loop structure.
The emission in the loop region is fainter in the images above 2keV
 and contains about half
of the counts in the soft image.
The low counts are insufficient to
extract detailed spectral information; however, 
the loop clearly shows a soft X-ray
signature.
There is also a faint emission beyond the loop extending
to $\sim$4'' south of the central source. 
This feature could be
associated with a diffuse emission with a low surface
brightness (total $\sim$20 counts).

The multi-component structure in CH Cyg was not detected in 
the Galloway and Sokoloski (2004) image because they applied
gaussian smoothing with
a FWHM of 2'' which
resulted in
degraded resolution.
In Figure 4 we compare our soft band high-angular resolution image
with
their result.
Although they did detect an extension of a few arcseconds from the central
source, their smoothed
image did not permit detection of the multi-component structures, including
the
loop-like structure
separated from the central source, which
are clearly seen in our adaptively smoothed images.

We further explored the reality of these low-count structures 
by applying a statistical deconvolution technique
 EMC2 (Expectation
through Markov Chain Monte Carlo) specifically applicable to low-counts Poisson data (Esch {\it et al.} 2004).
This technique uses a wavelet-like multiscale representation of the true image 
in order to achieve smoothing at all scales of resolution simultaneously.
Wavelet decomposition of the image allows the Poisson likelihood to be factored into 
separate parts, corresponding
to the wavelet basis.
Each of these factors in the likelihood can be reparametrized as a split of the intensity 
from the previous, coarser factor. 
A {\it prior} is assigned to these splits (which can be viewed as smoothing parameters), 
and then a model is fit using Markov Chain Monte Carlo (MCMC) methods.
In this way  small and large-scale structures in the image are captured, e.g. diffuse emission
as well as point sources and sharp features in the image.
This Bayesian model-based analysis allows assessment of the uncertainties 
in the reconstructed image.
Details of this technique and examples are described in Esch {\it et al.} 2004.
The EMC2 has been applied to various data sets including to Chandra observations
(e.g. Esch {\it et al.} 2004, Karovska {\it et al.} 2005).

We used EMC2 and the simulated PSF to perform deconvolution of
the CH Cyg soft X-ray image.
Figure 5 displays the EMC2 deconvolved image showing
complex multi-component structures 
with significance above 3 $\sigma$. The structures are
similar to those
resulting from adaptive smoothing, but with improved resolution.
In addition to the loop-like structure to the south of the central source, the central source appears to have a compact core
and an elongated structure extending to  $\sim$ 0.7'' at the P.A. of 330 deg.
A fainter structure is seen to the south at about same distance from
the source at P.A. of 150 deg.

The structures detected in the X-ray images of CH Cyg are very similar
to the structures detected in radio and optical observations obtained 
close in
time to the Chandra observations as described in the following
section.

\section{COMPARISON WITH HST AND VLA OBSERVATIONS}

Optical/UV and radio images of CH Cyg
obtained since  1996
show a complex circumbinary environment including ionized outflow,
and structures associated with
jet or outburst activity in the system (e.g. Corradi {\it et al.} 2001;
Eyres {\it et al.} 2002).

Although there are no observations contemporaneous with the 2001 Chandra
observation, we identified a set of VLA and HST observations that were carried out close in
time to the Chandra observations.
Radio maps of CH Cyg were obtained using the VLA at 5GHz and 8GHz on 
2000 November 9 about 4.5 months before the Chandra observation.
HST observations were carried out
on 1999 August 13 (in the optical, with WFPC2 ), 1.6 years 
before the Chandra observations.

The loops detected in the soft X-ray image  are
similar to the arc-like extensions
observed in the HST WFPC2 images (Eyres {\it et al.} 2002, and Crocker
{\it et al.} 2002).
The HST/WFPC2 images obtained
1.6 years before the Chandra observations 
are dominated by the
UV continuum, [OIII], H~$\alpha$ and H~$\beta$ lines. They
show extended emission
to the  north and to the south of the central region. 
Several structures, including an arc-like structure  1.2''
to the south
of the
central source, have been detected in
this extended emission (e.g. Eyres {\it et al.} 2002).
The loop-like structure is most clearly visible in the 502N [OIII]
image.

In Figure 5 we compare the deconvolved Chandra
soft X-ray image with the HST [OIII] 
image which has a similar resolution.

We note that the HST observation is the same as shown in Figure 1 
in Galloway and
Sokoloski (2004).
The extended [OIII] structures show 
three
distinct components: a central source, $\sim$0.5'' extensions to the north 
well as at $\sim$ PA160 deg to the south, and
a loop-like feature to the south at 1.2''.
The [OIII] loop-like feature fits 
within the loop in the Chandra soft image, as shown in Figure 5.
The south side of the [OIII] loop appears brighter as
seen in the X-ray image.

The VLA map shown in Figure 6 was made
from data taken on November 9, 2000 (VLA in the ``A'' configuration).  
The resolution is
$0.55" \times 0.39"$ with major axis position angle = $-82^o$, and
the
rms = 35 $\mu$Jy beam$^{-1}$.  This map shows a very similar
morphology when compared to the soft band Chandra image; it shows a
central structure extended at P.A.$\sim$ 330 degrees, as well a loop like structure
$\sim$1.5'' to the south.  The total flux density from the source is
7.2 mJy, with a peak surface brightness of 2.4 mJy beam$^{-1}$ at
19$^h$ 24$^m$ 33.03$^s$ +50$^o$ 14$'$ 27.1$''$ (J2000).  The loop
structure shows enhancement in the south side, similar to that of
the X-ray loop. The southern loop has an integrated radio flux of 1.6
mJy.  
The core source has an inverted spectrum of index $\alpha = +0.6$,
between 5 and 8 GHz, suggesting optically thick emission.  
We define $\alpha$ as $S_\nu \propto \nu^{\alpha}$.
The northern
extension has flat spectrum, $\alpha \sim 0$, suggesting optically thin
free-free emission.  The southern loop has a steep spectrum, $\alpha \sim
-0.8$, implying optically thin non-thermal emission.
The
minimum energy magnetic field in the southern loop based on the
non-thermal radio surface brightness is about 1mG, with a minimum
pressure of $6\times 10^{-8}$ dynes cm$^{-2}$.

In Figure 6 we display an overlay of the radio contours on 
the smoothed soft X-ray
image,
which has a similar resolution.  The radio and the X-ray loops are 
cospatial within the resolution limit of the VLA data of $\sim$0.4''.
We note that the resolution in the X-ray and radio images
is insufficient to determine if the regions of X-ray and radio emission
are separated.

\section{DISCUSSION}

Our analysis of the Chandra data shows that there are several 
spatially distinct
X-ray emission components in the CH Cyg system. 
The X-ray emission 
in the CH Cyg image
is dominated by an unresolved central core, however, there are several
extended components that contribute as well.
The emission from the central core
originates within a region of less then 50 AU, 
which is only few times larger
than the binary separation.
This is as close as we can get
to the central region of an interacting binary 
with today's X-ray imaging capabilities.

The hard emission from the central core, including in the iron line,
probably originates in
the inner region of the
accretion disk or in a boundary layer between the accretion disk and the
white dwarf.
The soft emission from the unresolved core may be attributed to the
accretion disk itself or to scattered hard X-rays from
the inner disk region by the outer disk or corona. The scattering
could be due to a non isotropically distributed 
 ionized photoionized medium similar to Seyfert 2 galaxies
(e.g. Wheatley and Kallman 2006).
The X-ray emission in CH Cyg core appears harder than in other symbiotics, 
eg. in R Aqr and Mira AB (Kellogg et al 2001, Karovska
et al 2005),  which have
relatively soft spectra, close to 1 kev.
Geometry of the surrounding material may play a role since part 
of the hard X-rays in these
systems may be absorbed or scattered 
in a different way.

In addition to the unresolved core, the soft X-ray image shows 
 a second component in the central region, an extended
structure to $\sim$ 150 AU at 330/150 degrees.
This structure was also
detected in the radio, and HST O[III] images.
The emission could be due to a
shocked gas formed in the region of collision between the  
fast wind from a white
dwarf with the red giant wind. 
However, since the orientation of this structure 
is similar to that of radio and optical
jets observed  in
the past, and to the general orientation of the loop
structures further to the south (e.g. Eyres {\it et al.} 2002, Corradi {\it et al.} 2001),
 we believe that the structure is more likely  
associated with a recent jet activity, e.g., during
the 1998-2000 active phase. 
The emission could
result from jet interaction with the close 
circumbinary material including circumstellar dust shells surrounding
the giant.
In fact, Mid-IR photometry does show the presence of a significant dust in the vicinity of the central region of CH Cyg
(Teranova and Shenavrin 2004). Furthermore, high-angular resolution Mid-IR
imaging of CH Cyg
detected a dusty
circumstellar feature $\sim$0.7'' to the North of central source,
similar to the extension of the feature detected in Chandra, Radio and
HST images (Biller {\it et al.} 2006).  Follow up observations of this
extended X-ray structure are needed to
determine its nature and kinematics.

Our imaging shows that although the X-ray emission from
the 
central region
dominates the soft emission in CH Cyg,
a significant fraction ($\sim$20$\%$)
originates further away,
from the loop at $\sim$ 400 AU and from the extended diffuse emission beyond.
The loop
is likely associated with the 1997 radio outburst which was followed
by jet activity (Karovska {\it et al.}
1998a, Karovska {\it et al.}
1998b).
The 1997 radio outburst in CH Cyg was detected using the VLA at several frequencies (Karovska et al
1998a).
Nearly contemporaneous
continuum measurements at millimeter and submillimeter wavelengths
with the IRAM
interferometer and with SCUBA on the James Clerk Maxwell Telescope
also showed evidence for a  strong radio outburst 
(Mikolajewska {\it et al.} 1998). 
The 1997 radio
outburst occurred following 
an unprecedented drop in the V magnitude in 1996 (Karovska et al
1998a), similar to a
previous event in 1984, 
following which a powerful jet was formed (Taylor {\it
et al.} 1986).
In early 1998 we detected a radio jet
using the VLA in its highest resolution observing mode (Karovska {\it
et al.} 1998b). 
At that time, the radio maps detected two structures with a roughly
North-South orientation, with the southern structure 
separated from the central source by
about 0.35''. 
The VLA observations of the loop carried out later on in 2000
show that the structure has
expanded to a distance close to 1.5'', or $\sim$ 375AU.
This corresponds to an expansion velocity of $\sim$ 400 km/s.

Assuming that the loops we detected in the X-rays and radio, and in [OIII], are a
consequence of the same outburst event,
the [OIII] loop would have had to expand at a velocity of
several hundred km/s to reach the position seen in the Chandra
images 1.6 years later. 
It would also have taken a few years for ejecta from the central region with
an expansion velocity of few
hundred km/s to reach the
apex of the [OIII] loop, and then to expand to the region 
of the radio and X-ray loops (assuming that the ejecta
 have not been slowed down since the outburst).
This is
in agreement with the apparent expansion velocity derived from the
1998 VLA observations.

Kinematic studies of outflows in CH Cyg do show velocities of the expanding
material on the order of few hundred km/s (Corradi {\it et al.} 2001)
which is in agreement with the velocity of expansion that we calculated.
On the other hand, spectroscopic observations also show evidence for
mass 
ejections in CH
Cyg that indicate
much higher velocities, well over 1000 km/s (e.g. Skopal {\it et al.} 2002).
Skopal {\it et al.} (2002) suggested that this difference could be due to a
significant deceleration of the ejecta due to interaction with the
circumstellar medium, which would be also in agreement with our observations.
On the other hand, if the ejecta have not
been slowed down and are moving with a speed over 1000 km/s, there remains a 
possibility that the X-ray and radio emission in the region of the
loop may be a consequence of an outburst during the more recent 1998-2000 active phase.

The X-ray emission from the loop is consistent
with a shock generated by ejecta 
colliding with the
surrounding pre-existing circumstellar material.
The X-ray emission beyond the loop could be associated with
continuous interaction of
the expanding wind with the pre-existing circumbinary material,
or it could result from more intermittent shocks associated with 
a pulsed jet (e.g., Stute 2006).  The non-thermal spectral index of the
radio emission could be explained by particle acceleration at the 
shock.  

The small number of counts in the loop do not permit us to fit for
the temperature and absorbing column of the emitting gas.  If we make
the most favorable assumptions given the 60 counts detected and
the hardness ratio, the temperature is about 1 keV and the luminosity 
is about $5 \times 10^{28}~\rm erg~s^{-1}$.  For a 400 AU sphere
and an X-ray emissivity of $2 \times 10^{-23}~\rm erg~cm^3~s^{-1}$
(e.g. Raymond, Cox \& Smith 1976), the density is about 50 $\rm cm^{-3}$.
This estimate is probably on the low side, given that the spherical 
volume is not filled and that absorption will reduce the observed
count rate.  Nevertheless, it implies a pressure about 2.5 times the
minimum magnetic pressure estimated from the radio emission.  The mass of the
X-ray emitting gas is about $10^{26}$ g, or 0.1, 0.05 and 1.0 times the
masses of jet components A, B and C measured by Taylor et al. (1986),
respectively.  Given that the velocity ellipses measured by Corradi
et al. (2001) suggest shock speeds of about 100 km/s, while the
X-rays require shock speeds near 1000 km/s, the simplest interpretation
is that the jet is underdense by a factor of 100 compared to the
ambient gas.  The slower shock in the ambient gas produces Balmer line,
[O II] and [O III] emission by a combination of shock heating and
photoionization of the compressed gas. Because the gas cools quickly
the compression is very strong.  The fast shock produces X-rays,
but a shock speed closer to 2000 $\rm km~s^{-1}$ than 1000  $\rm km~s^{-1}$
may be required, since electron temperatures are generally well below
proton temperatures in such fast shocks (e.g. Rakowski 2006).
The radio emission could come from either shock, but the
strong magnetic field suggests that it originates in gas that
has been strongly compressed by the slower shock. If the first shock
generates the radio emission, a significant magnetic field, on the
order of 25 $\mu$G, would be required in the jet.

A high speed (precessing) jet may also clear a cavity in the pre-existing
circumstellar material (e.g. Ybarra {\it et al.} 2006).  The X-ray and the radio emission in that case
originate from a loop-like structure delineating a shock
region in the boundary between the walls of the cavity and the
surrounding material.  
A similar scenario has been suggested for the loop-like
structure observed previously further away ($\sim$5'') from the central source
in the UV O[II] HST images of CH Cyg(Corradi {\it et al.} 2001).

\section{CONCLUSIONS}

Our study of the CH Cyg close circumbinary environment using a 
combination of multi-wavelength 
X-ray, radio and optical high-angular resolution observations
identified several  distinct components in this system. These
include a compact central
source, and several extended structures likely associated with jet
activity.
These results are important since
CH Cyg is only the second symbiotic system after R Aqr (Kellogg {\it et
al.} 2001)
with jet activity detected at X-ray wavelengths, and 
the Chandra image combined with the VLA and HST data
provides the closest view of 
the region where jets are being formed and interacting with the
surrounding material
as they propagate
through the circumbinary environment.
Furthermore, our analysis of the Chandra data resolves the ambiguity as to whether there is only one
(Wheatley and Kallman 2006), or multiple components (e.g. Ezuka {\it et
al. 1998}, Mukai {\it et al.} 2006) contributing to
the X-ray  emission from this system.

Continuing multiwavelength monitoring of CH Cyg is necessary
to determine the characteristics of the multi-component
structures and their evolution as they
interact  with the surrounding
circumbinary environment. It is especially important to obtain longer
exposure X-ray observations of this system which will increase the
signal-to-noise in the images and allow deriving
spectral
characteristics of the emissions from individual components.

\acknowledgments

Janet Mattei passed away before the completion of this article, and her coauthors respectfully
dedicate it to her memory.
We are grateful to W. Hack for providing the drizzled HST image of
CH Cyg, and to AAVSO for providing the CH Cyg light curve.
We thank Y. Butt for useful discussions.
We would like to thank the anonymous referee
for helpful comments and suggestions.
MK is a member of the Chandra X-ray Center,
which is operated by the Smithsonian Astrophysical Observatory under
NASA Contract NAS8-03060. CC would like to acknowledge support from
the Max-Planck Society and the Alexander von Humboldt Foundation
through the Max-Planck-Forschungspreis 2005.


\begin{figure}
\epsscale{.85}
\plotone{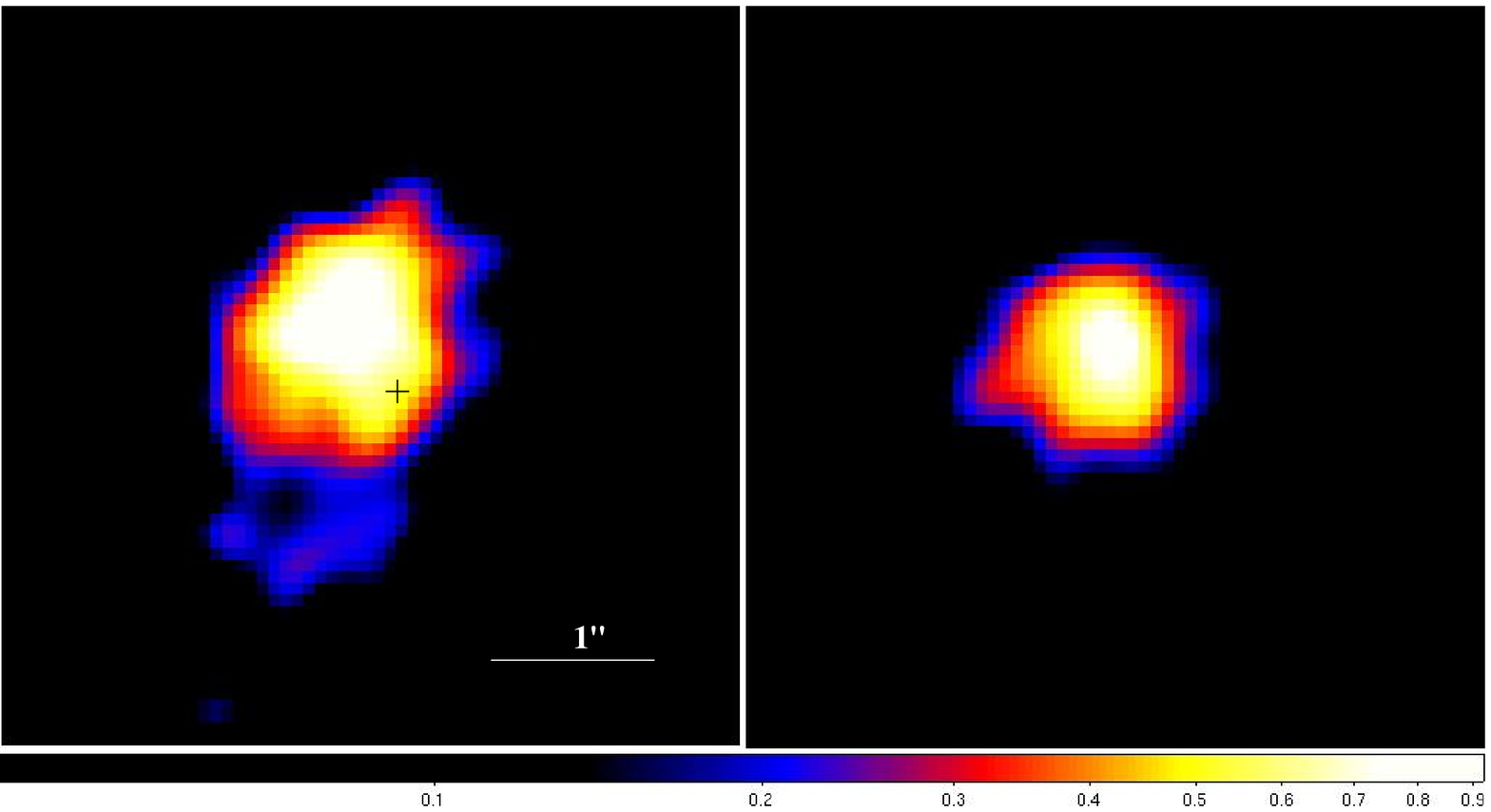}
\caption{Adaptively smoothed
ACIS-S images of 
CH Cyg in two spectral bands: 
left - soft band (0.2-2 keV) image,
right -  hard band (2-6 keV) image. North is up, East
is to the left.}
\end{figure}

\begin{figure}
\epsscale{.85}
\plotone{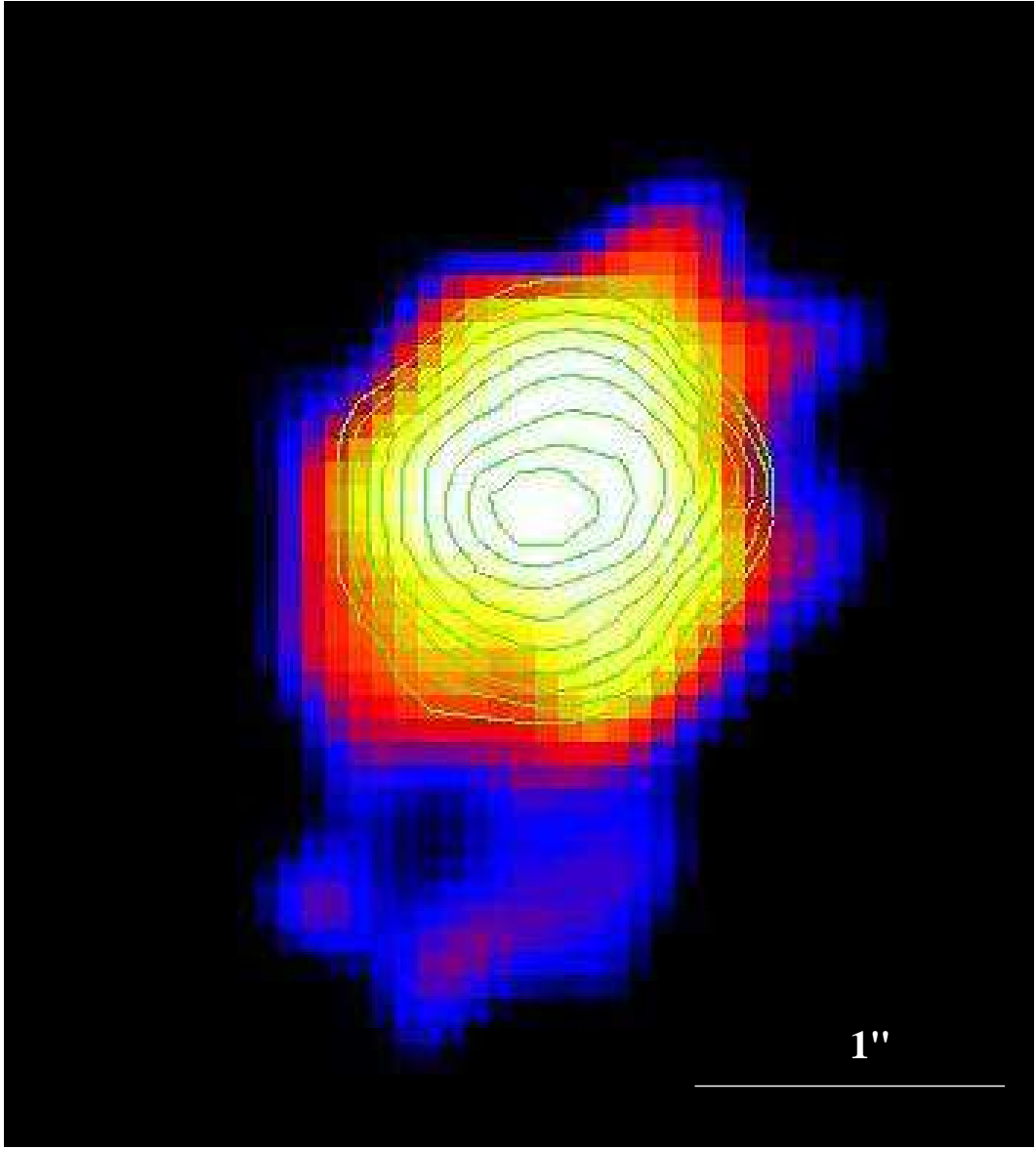}
\caption{
CH Cyg soft image with overlayed contours of the simulated PSF showing the loop at 1.5'' from the
central source. The central source appears elongated in the N-W direction.
North is up, East
is to the left.}
\end{figure}

\begin{figure}
\epsscale{.85}
\plotone{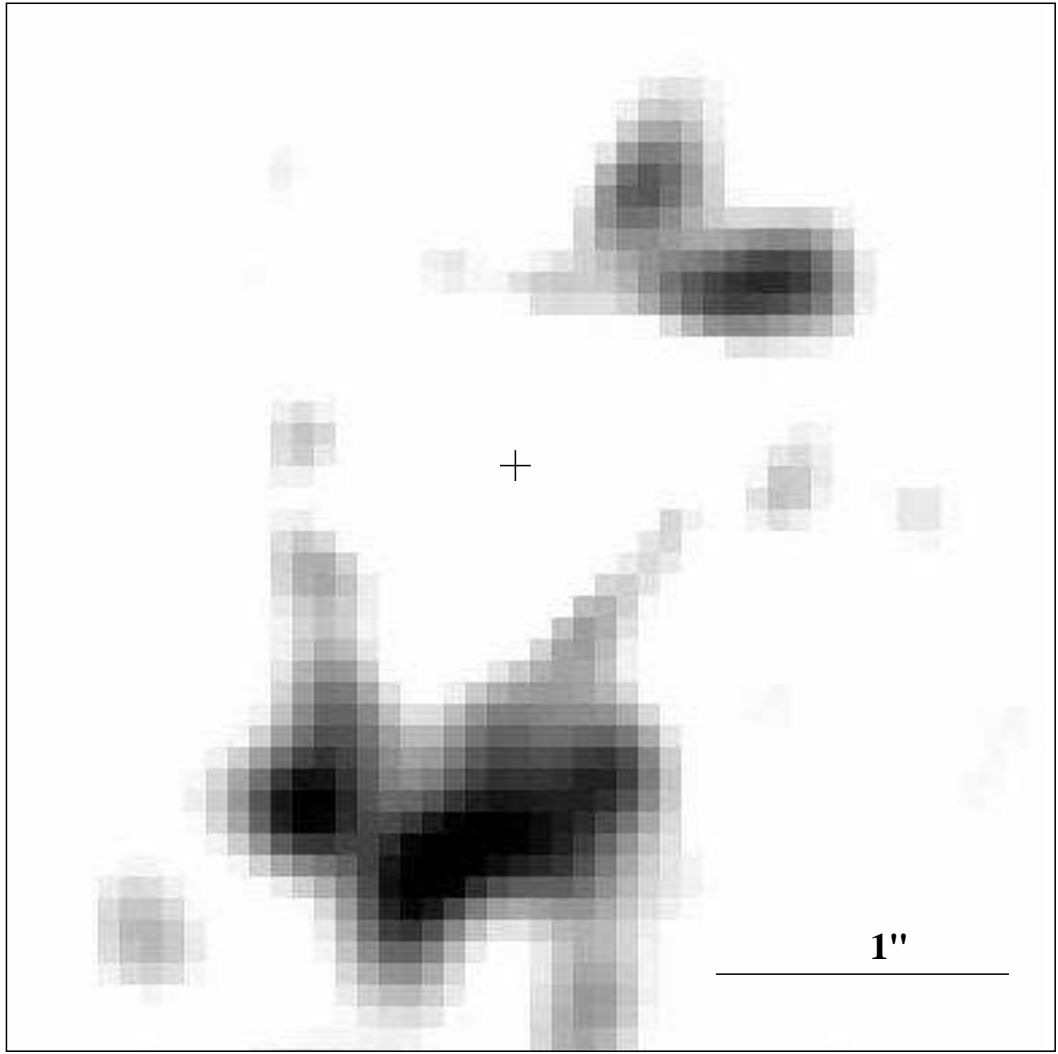}
\caption{Soft image residuals after subtraction of the PSF show the
loop-like structure to the south and extension toward the N-W at
$\sim$330 degrees}
\end{figure}

\begin{figure}
\epsscale{.85}
\plotone{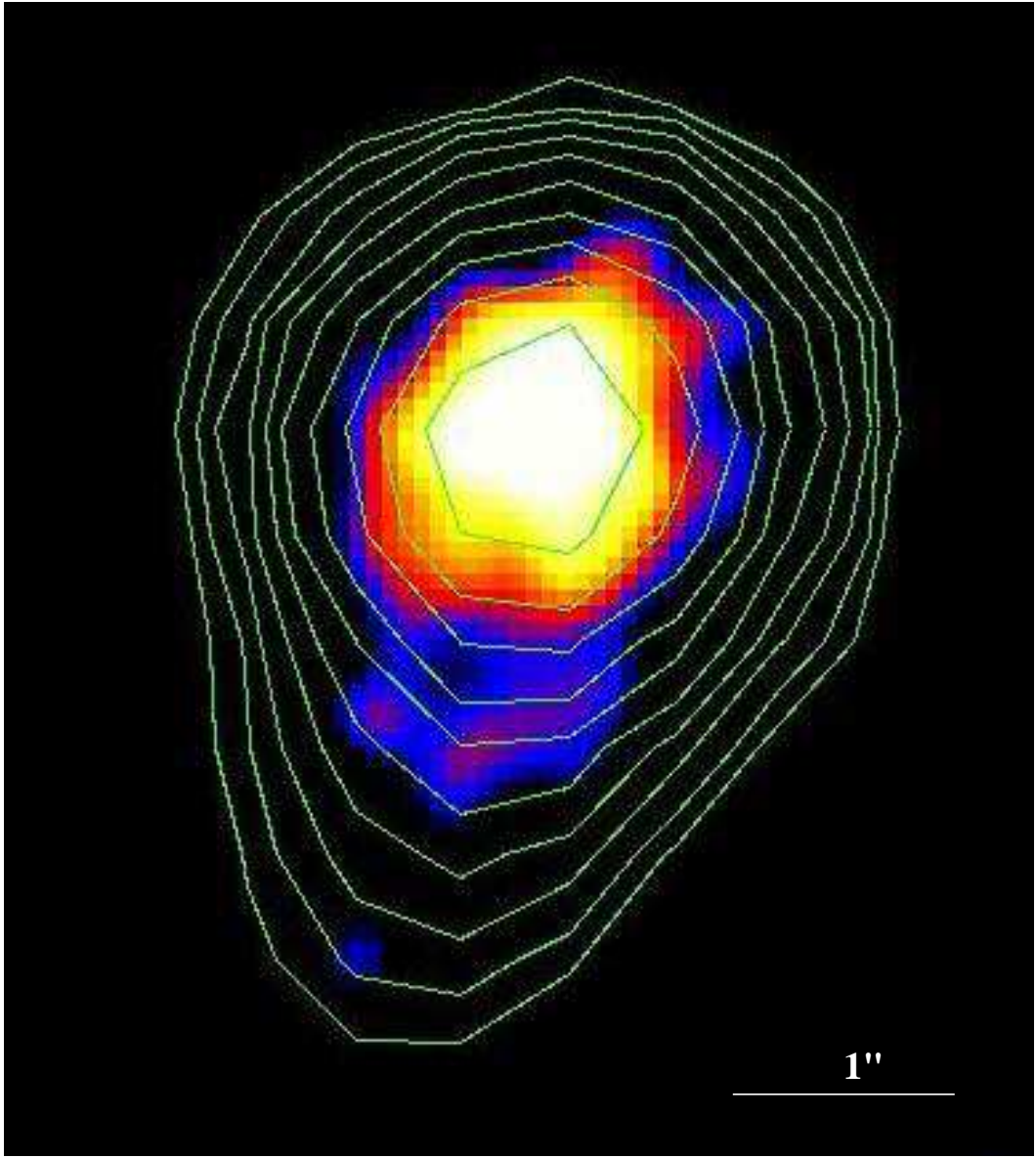}
\caption{Comparison of the Chandra soft band adaptively smoothed image
showing the central source and the loop with
the Galloway and Sokoloski (2004) image (contours). The latter was smoothed 
using a gaussian with FWHM of 2''.}
\end{figure}

\begin{figure}
\scalebox{0.998}{\rotatebox{0}{\includegraphics{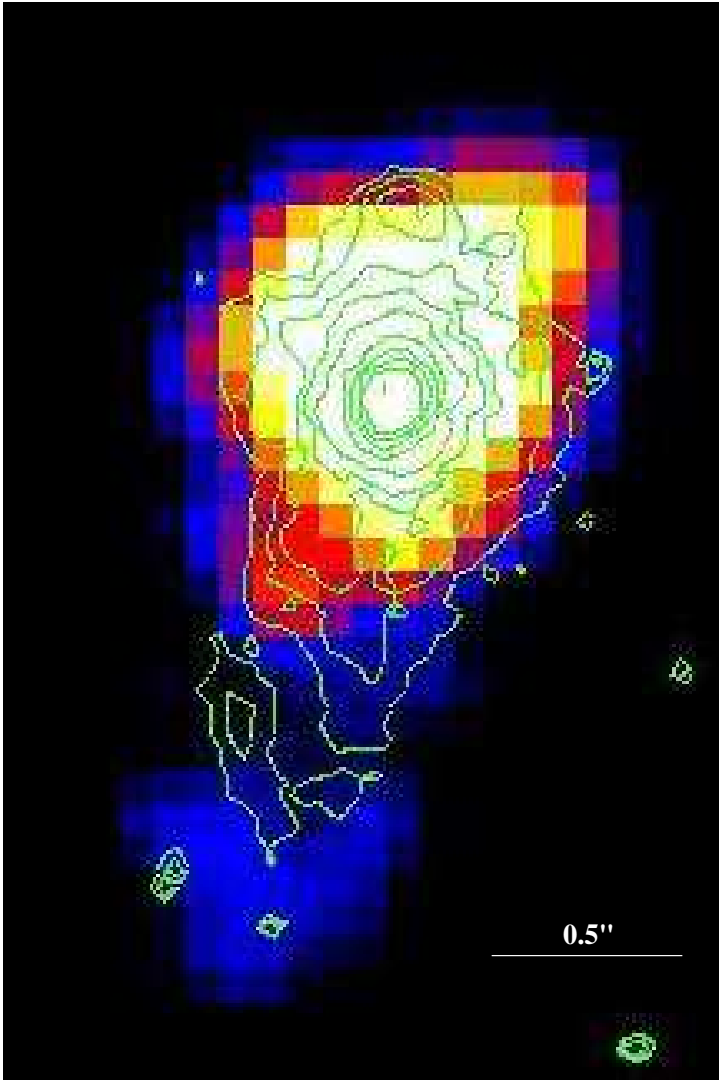}}}
\epsscale{.85}
\caption{Chandra soft band image of CH Cyg deconvolved using simulated
PSF. The image shows
a complex multi-component structures 
 similar to the structures detected 
in the adaptively smoothed image shown in Figure 2 but with 
improved resolution.
In addition to the loop-like structure to the south of the central
source, an
elongated structure is detected at $\sim$ 0.7'' from the central source 
in the direction of 330/150 degrees. Overlayed are contours of an OIII
HST image obtained 1.6 years before the Chandra image showing similar 
morphology.}
\end{figure}

\begin{figure}
\epsscale{.85}
\plotone{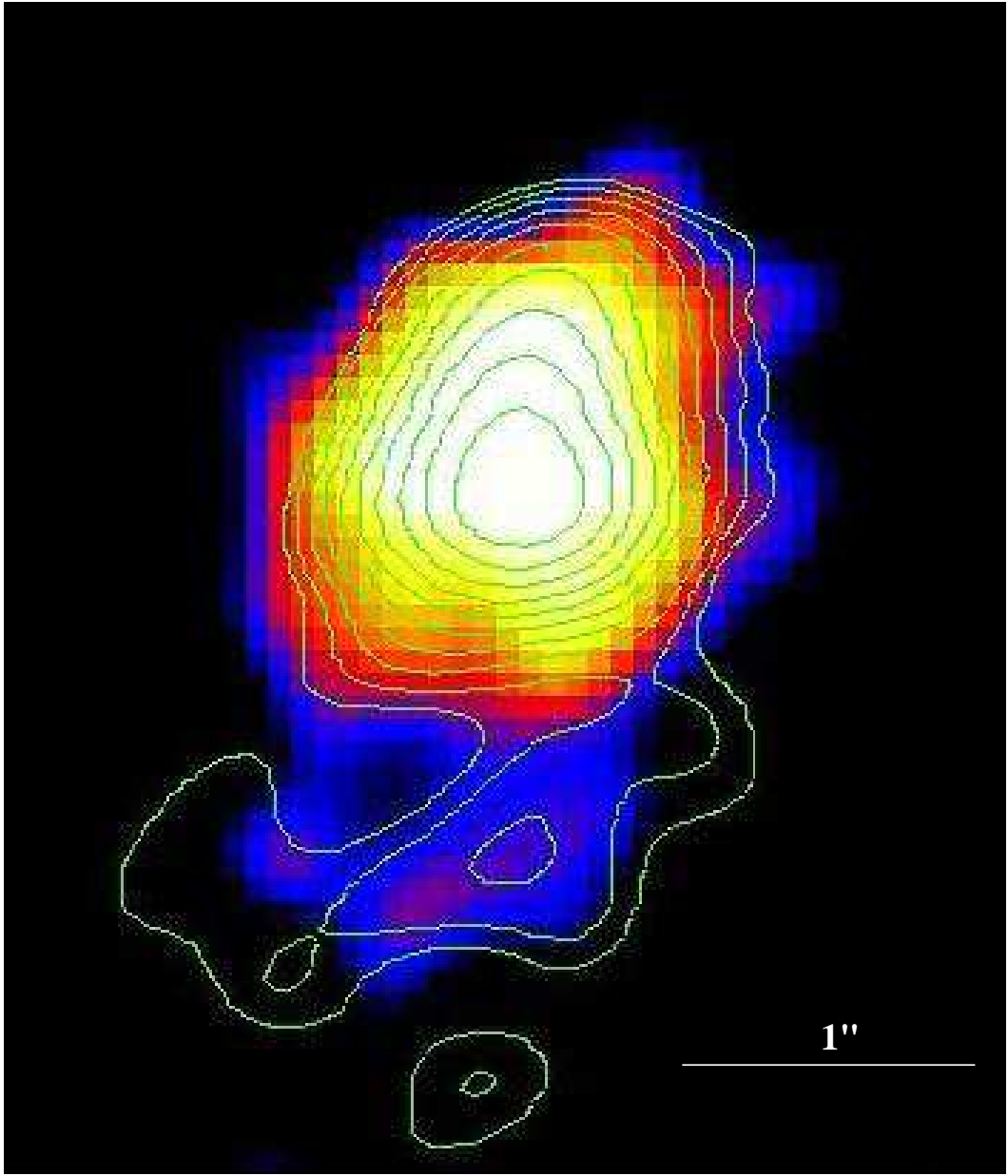}
\caption{Overlay of 5GHz radio image contours on the soft X-ray
smoothed image showing that the X-ray and radio loops are cospatial
within the resolution of 0.3-0.4''.
The radio image was obtained 4.5 months before the Chandra image.}
\end{figure}


\begin{thebibliography}{}

\bibitem{Biller} Biller {\it et al.} 2006, astro-ph 0604336

\bibitem{Burrows2000} Burrows, D.N. {\it etb al.} 2000, ApJ, 543,149


\bibitem[\protect\citeauthoryear{Chugai and Yungelson}{2004}]{Chugai2004} 
Chugai, N.N., and Yungelson, L.R.,, 2004, Astron. Lett., 30, 65.

\bibitem[\protect\citeauthoryear{Corradi {\it et al.}}{2001}]{Corradi2001} 
Corradi R.L.M. et al., 2001, ApJ, 560, 912.

\bibitem{Crocker2002} Crocker, M.M., Davis, R.J., Spencer, R.E.,
Eyres, S.P.S., Bode, M.F., Skopal, A., 2002, MNRAS, 335, 1100

\bibitem{Ebeling 2006} Ebeling, H.; White, D. A.; Rangarajan,
F. V. N. 2006, MNRAS, 368, 65

\bibitem{Eyres2002} Eyres, S. P. S., Bode, M. F., Skopal, A., Crocker,
M. M.; Davis, R. J., Taylor, A. R., Teodorani, M., Errico, L.,
Vittone, A. A., Elkin, V. G. 2002, MNRAS, 335, 526

\bibitem{Esch2004} Esch, D.N., Connors, A., Karovska, M., van Dyk,
D.A., 2004, ApJ, 610, 1213

\bibitem{Ezuka 1998} Ezuka {\it et al.} 1998, ApJ, 499, 388

\bibitem{GallowaySokoloski2004} Galloway, D.K. Sokoloski, J.L. 2004,
ApJ Letters, 613, 61

\bibitem{Hinkle1993} Hinkle, K.H. et al 1993, AJ, 105, 1074.

\bibitem{Karovska1998} Karovska, M., Carilli, C.,  and Mattei, J.,
1998a,
IAU Circ. No. 6970.

\bibitem{Karovska1998} Karovska, M., Carilli, C.,  and Mattei, J., 1998b, JASVO,
26, 97.

\bibitem{Karovska1992} Karovska, M. and
Mattei, J., 1992, $JAAVSO$, {\bf 21}, 23

\bibitem[]{Karovska2006} Karovska, M., Schlegel, E., Hack, W., Raymond, J. C., \&
Wood, B. E. 2005, ApJ, 623, L137

\bibitem[]{Karovska2006} Karovska, M. 2003, Chandra Newsletter, Vol. 10, p.20-21

\bibitem{Kellogg} Kellogg, E., Pedelty, J., and Lyon, R. 2001, ApJ, 563,
L151.


\bibitem{LeahyTaylor} Leahy, D.A. \& Taylor, A.R. 1987,
AA, 176, 262

\bibitem{LeahyVolk} Leahy, D.A. \& Volk,  1957,
ApJ, 440, 847

\bibitem{Mikolajewska1998} Mikolajewska {\it et al.} 1998, $IAU Circ.$ No. 6968

\bibitem{Mukai2006} Mukai, K., Ishida, M., Kilbourne, C., Mori, H.,
Terada, Y., Chan, K-W, 2006, PASJ, in press 

\bibitem{Park2002} Park, S. {\it etb al.} 2002, ApJ, 567, 314


\bibitem{Perryman1997} Perryman {\it etb al.} 1997, A\&A, 176, 262

\bibitem{rakowski} Rakowski, C.E. 2006, Ad. Sp. Res. 35, 6, 1017

\bibitem{Raymondetal} Raymond, J.C., Cox, D.P. \& Smith, B.W. 1976, ApJ, 204, 290 

\bibitem{Skopal2002} Skopal, A., Bode, M.F., Crocker, M.M., Drechsel, H., Eyres,S.P.S., Komzík,R. 2002, MNRAS, 335, 1109

\bibitem{Sokoloski and Kenyon} Sokoloski, J.L. and Kenyon, S. 2003, ApJ,
584, 1021)

\bibitem{Stute} Stute, M. 2006, A\&A, 450, 645

\bibitem{Taylor} Taylor, A.R., Seaquist, E.R., \& Mattei, J.A. 1986,
Nature, 319, 38

\bibitem{Teranova} Teranova, O.G., \& Shenavrin, V. I. 2004,
Astronomy Reports, 48, 813


\bibitem{Weisskopf} Weisskopf, M.C. {\it et al.} 2002, PASP, 114, 1

\bibitem{Wheatley2006} Wheatley, P.J, \& Kalmman T.R. 2006, MNRAS, 372,
1606

\bibitem{Ybarra} Ybarra {\it et al.} 2006, astro-ph/0608162

\end{thebibliography}
\end{document}